# Temperature and Face Dependent Copper –Graphene Interactions

Sara D. Costa, Johan Ek Weis, Otakar Frank, Martin Kalbac*

J. Heyrovský Institute of Physical Chemistry, Academy of Sciences of the Czech Republic, v.v.i., Dolejškova 3, CZ-18223 Prague 8, Czech Republic.

The interaction between graphene and metals represents an important issue for the large-area preparation of graphene, graphene transfer and the contact quality in graphene devices. We demonstrate a simple method for estimating and manipulating the level of interaction between graphene and copper single crystals through heat treatment, at temperatures from 298K to 1073K. We performed an in-situ Raman spectroscopy showing Cu face-specific behavior of the overlying graphene during the heat treatment. On Cu(111) the interaction is consistent with theoretical predictions and remains stable, whereas on Cu(100) and Cu(110), the initially very weak interaction and charge transfer can be tuned by heating. Our results also suggest that graphene grown on Cu(100) and Cu(110) is detached from the copper substrate, thereby possibly enabling an easier graphene transfer process as compared to Cu (111).

1. Introduction

Graphene is an outstanding material and one of the greatest challenges is to obtain large and uniform single layers of graphene, that are easily transferable. Chemical vapor deposition (CVD) is one of the mostly used techniques to obtain large graphene layers, containing single crystalline graphene [1, 2] or polycrystalline graphene [3]. CVD growth of graphene involves the presence of a catalyst, a carbon source and high temperatures. Thin copper foils are often used as catalysts for graphene growth because of their relatively low cost, availability and high yields of monolayer graphene. Moreover, the copper foil can be removed after the growth process, in order to transfer the graphene layer onto different substrates [4, 5]. However, as the copper foils are polycrystalline, the resultant products can vary due to the presence of grain boundaries and other inhomogeneities

---

* Corresponding author. Tel: +420 266 053 804. E-mail: martin.kalbac@jh-inst.cas.cz (Martin Kalbac)

[6]. A single crystal represents the thermodynamically most stable form of metal, provided it is clean and free from mechanical strain [7]. Hence, the investigation of copper single crystals as catalysts for graphene growth is of high relevance. Copper single crystal properties, such as electrical resistivity, thermal conductivity and thermal expansion as a function of temperature have been investigated since the beginning of the last century [7, 8]. The results have shown an increase of the electrical resistivity and of the thermal expansion coefficient with temperature. The thermal expansion of copper is not linear for the whole range of temperatures (10-1200 K) [9], with a value of 16.65 x $10^{-6}$ $K^{-1}$ around room temperature, and 22.3 x $10^{-6}$ $K^{-1}$ at 1000K [10].

The influence of the surface orientation of copper single crystals on the number of layers of graphene, size and orientation of the grown grains has been reported by several authors [6, 11, 12]. Murdock *et al.* [6], reported on graphene domains with grain shapes and orientation changing according to the crystallographic orientation of the copper substrate. For instance, four-lobed grains were grown on Cu(100) and Cu(110), while elongated hexagonal grains were obtained for Cu(111) substrates [6]. Wood *et al.* [12] investigated graphene grown on copper foils with different lattice orientations, observing monolayer and bilayer graphene for Cu(111) and Cu(100), respectively. The difference in the number of layers was justified by the growth rates, which were faster for Cu(111) and slower for Cu(100) [12]. Different doping levels and strain on graphene grown on Cu(100), Cu(110) and Cu(111) have also been reported by Frank *et al.* [11]. It was found that graphene grown on Cu(111) was *n*-doped, while graphene samples on Cu(110) and Cu (100) were not doped or partially doped, respectively [11]. Additionally, it was also reported that graphene layers were compressed, from 0.1 to 0.3%, on all the investigated copper orientations [11].

Most of the studies found in the literature have used Raman spectroscopy due to its non-destructibility and high sensitivity to various parameters of the materials, such as doping [13-15], strain [16, 17], and thermal heating properties [18, 19].

The thermal properties of graphene have been studied before, utilizing either local heating (using high laser powers during Raman spectroscopy measurements [19]), or heating cells in inert atmosphere [18]. Despite the differences, both types of experiments show that the increase of temperature leads to a downshift in the frequency of the Raman modes, and vice-versa. The observed frequency shifts vary depending on the heating source (laser or heating cell), the number of layers, and the adhesion between the graphene layers and their substrate [18-20]. The transfer of graphene onto substrates other than copper is essential and it represents one of the most challenging steps towards the preparation of clean and homogeneous graphene layers. Great efforts have been employed to improve the graphene transfer process from copper, using a variety of techniques, such as PMMA mediated transfer [21], both using wet and dry processes [5], bubbling transfer [22], hot pressing [4], roll to roll techniques [3], and by high electrostatic potentials [23]. Recently, new methods involving peel-off of graphene from copper [24, 25], instead of chemical etching of the copper foil, which often leave remnants of copper on the sample, have been developed.

The level of detachment of graphene from the copper surface is crucial to provide a better transfer, however, this parameter is difficult to quantify as the most common copper foils are polycrystalline. The detachment of graphene from the copper is directly related to the induced doping on the graphene layer and the investigation of graphene grown on copper with a single orientation provides a possible approach to estimate the degree of detachment of graphene layers. Furthermore, the interactions between metals and graphene are of a general importance because of their effects on electrical contacts in graphene based electronic devices [26, 27].

The study of the thermal properties of the graphene layers grown on single crystals is also of high relevance since high temperatures are employed during the CVD process, which can influence the properties of the final product. The investigation of the growth mechanism of graphene on copper single crystals has been reported before, but a systematic study on heating/cooling effects on this material has not been reported yet, as it is a challenging experiment.

This manuscript reports on how copper induced doping of graphene layers can be used as a guideline to estimate the interaction between graphene and copper, and the influence of the temperature and copper surface orientation. An *in-situ* Raman spectroscopy study on single layer (1-LG) graphene grown on copper single crystals with three different lattice orientations - (100), (110) and (111) – was performed. In order to evaluate the temperature effects on the graphene layers, the samples were subjected to heating and cooling cycles in an inert atmosphere, with simultaneous acquisition of Raman spectra. The results show that graphene grown on Cu(111) is doped similarly before and after heating, as the contact between the graphene and the copper is better than in other orientations, due to their closely matched hexagonal lattices. Contrarily, graphene on Cu(100) and Cu(110) becomes doped with the increase of temperature, as the contact between the graphene layer and the copper surface increases, raising the interaction between the two materials. The lack of doping in ambient conditions suggests that graphene on Cu(100) and Cu(110) is detached from the copper, due to their mismatched lattices, thereby possibly inducing an easier removal of the graphene layer during the transfer process.

2. **Experimental section**

*Graphene growth:* The graphene layers were grown by CVD on Cu single crystal substrates with lattice orientations of (100), (110) and (111) (MaTecK, diameter of 5mm, purity 99.9999%, orientation accuracy <0.1º). Firstly, the substrates were heated to 1273 K and annealed for 20 min under flowing $H_2$ (50 sccm). Methane was then introduced for another 20 minutes. Finally, the samples were cooled down from 1273 to 273 K under an $H_2$ flow. The pressure was kept at 0.35

Torr during the whole growth. In order to avoid the effect of possible temperature gradients or other differences in the growth conditions, all three Cu single crystals were placed side by side perpendicularly to the length of the quartz tube, being subjected to the growth procedure simultaneously. The order, in which the copper crystals were placed in the furnace, was changed for different growths, and no difference was observed in the results.

*Heating experiments:* In order to investigate the thermal properties, the samples were placed, individually, in a commercial heating cell (Linkam). The cell was placed under the microscope of the Raman setup and the measurements were acquired through an optical window located right above the sample. The heating cell was connected to a pump and a gas source. Before the experiment the cell was pumped out to remove oxygen and during the experiment a flow of He gas was supplied in order to keep the chamber free from oxygen and to ensure good temperature control of the sample. The heating process was carried out in two heating/cooling cycles, from room temperature to 1173 K, pausing every 100K to acquire Raman spectra.

*Raman Spectroscopy: In situ* Raman spectra were acquired every 100 K of the heating cycles. Raman measurements were performed using a LabRAM HR Raman spectrometer (Horiba Jobin-Yvon) and an Ar/Kr laser (2.54 eV/488 nm, Coherent). A 50x objective was used, providing a laser spot of about 1 μm. The laser power was 1 mW outside of the heating chamber, corresponding to about 0.7 mW at the sample after going through a sapphire window and an aperture in a heat radiation shield.

In order to ensure the reproducibility of the data, the experiments were repeated three times and similar results were obtained in all runs.

3. **Results**

Copper crystals are represented by cubic unit cells, and have a face centred cubic (*fcc*) structure. The surface of the copper single crystal can present different lattice orientations depending on the plane direction as shown in Figure 1a, which influences the properties of the graphene grown on it. For instance, carbon hexagons are preferentially located above Cu atoms, at low density sites, which makes Cu(111) a favored orientation due its hexagonal configuration, leading to a reduction of lattice mismatch between the graphene and the copper [6].

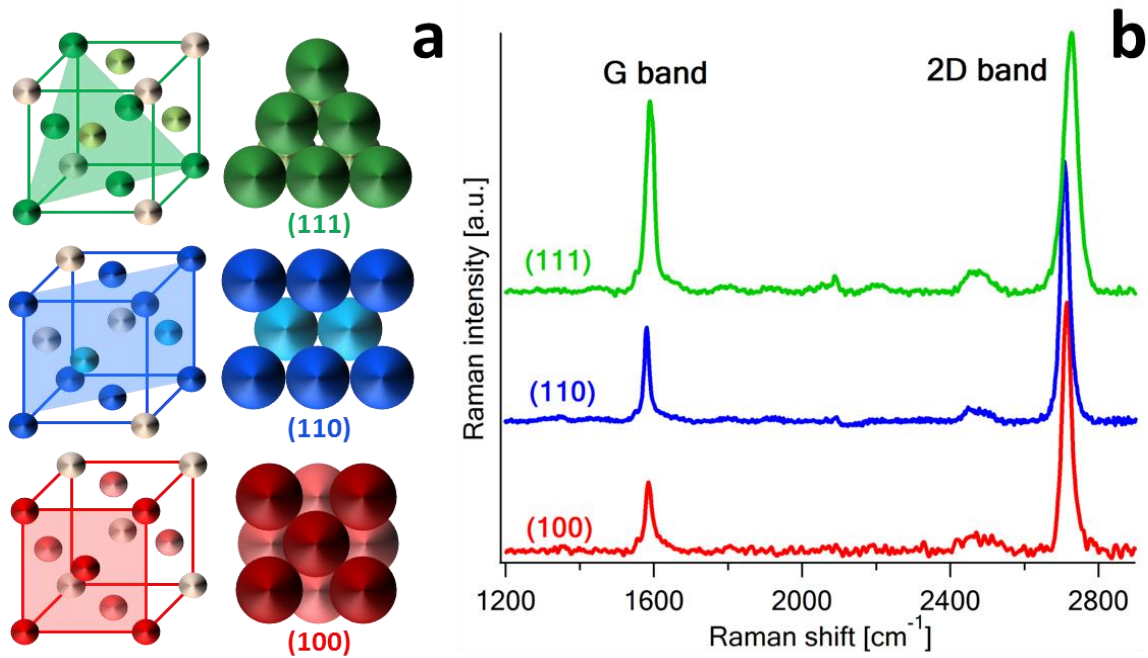

**Figure 1** (a) Scheme of plane and atom arrangements for each crystal surface: (111) – green, (110) – blue, and (100) – red. (b) Raman spectra of 1-LG graphene grown on copper single crystal substrates (at 2.54 eV), representing the main features: G band and 2D band. Each spectrum is an average of 30 spectra acquired on each sample. The spectra were normalised to the 2D band and are offset for clarity.

The graphene samples were analyzed via Raman spectroscopy, and the spectra indicate the presence of monolayer graphene (1-LG) (Figure 1b). The sample grown on Cu(111) shows a smaller 2D/G ratio, if compared to graphene grown on Cu(100) and Cu(110). Raman spectra with less intense 2D bands have been previously assigned to monolayer graphene doped by copper [11].

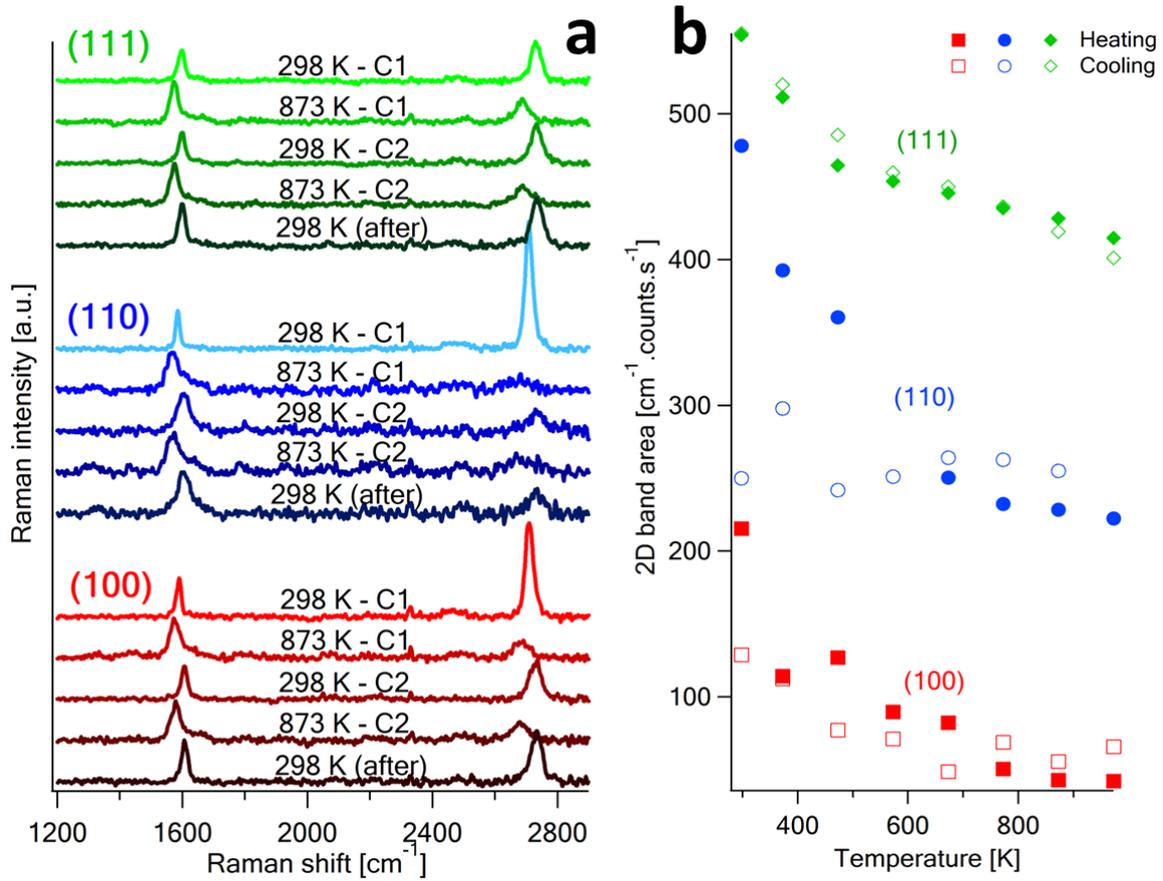

**Figure 2** (a) In-situ Raman spectra, at 2.54 eV, of 1-LG on copper single crystals at 298 K (room temperature) and 873 K, for the heating processes of cycle 1 (C1) and cycle 2 (C2), and 298 K after the thermal treatment. The spectra were normalised to the G band intensity and are offset for clarity. (b) 2D band area versus the temperature for heating and cooling of cycle 1.

In order to investigate graphene-substrate interactions for the graphene samples grown on the different copper single crystals, two heating/cooling cycles were carried out, while Raman spectra were acquired. Figure 2a shows representative spectra at room temperature (298 K) and at 873 K for the different samples, both for cycle 1 (C1) and cycle 2 (C2). For graphene grown on Cu(111) the intensity of the 2D band decreases at high temperatures, but its initial values are recovered when the sample is cooled. A similar behavior is observed for graphene grown on Cu(100), however, the intensity of the 2D band after cooling is only partially recovered, when kept under inert atmosphere. For graphene grown on Cu(110) the 2D band also decreases upon heating, however, it is not recovered after cooling. The decrease of the intensity of the 2D band upon heating can be caused by: 1) temperature enhanced effects of strain and doping [13, 28], which are reversible effects upon cooling, and/or 2) metal induced doping [15, 29, 30], which is irreversible in an inert atmosphere. The decrease of the 2D band intensity at high temperatures can be a consequence of charge transfer between the graphene layers and impurities not removed by the heating, affecting the electron-electron scattering process in graphene [13]. Additionally, the decrease in intensity of the 2D band has also been observed as a consequence of doping [13, 15], which in the presented case could reflect an improved contact between the Cu surface and the graphene layer, caused by the high temperatures. However, the initial conditions of the interactions between graphene and the differently-oriented Cu faces differ [11]. The lattices of graphene and single crystals with Cu(100) and Cu(110) orientations are highly mismatched [31], which possibly allows intercalation of molecules present in the atmosphere between the metal surface and the graphene. On the other hand, the contact between the graphene layer and the Cu(111) substrate is better due to the matching hexagonal lattices [31], and its affinity is not further improved at high temperatures, as evidenced in the unchanged 2D band present in the Raman spectra at room temperature after the heating/cooling cycle. Figure 2b shows the area of the 2D band (A(2D)) plotted for all the temperatures during heating and cooling of cycle 1. The A(2D) for graphene grown on Cu(111) shows a similar behavior for both heating and cooling processes, while for samples grown on Cu(100) and Cu(110), the A(2D) at room temperature after heating and cooling is lower than at the beginning of cycle 1. The biggest difference between A(2D) before and after cycle 1 is observed for the Cu(110) sample, which is most likely related to the different affinities of the graphene and copper crystal lattices. The values of A(2D) at high temperatures differ for each copper substrate, suggesting that different levels of doping are reached by each sample. It should be noted that in a similar experiment performed on CVD graphene [20] or exfoliated graphene [32, 33] on $Si/SiO_2$ substrates, the observed 2D band intensity decrease was also observed. Nevertheless, the doping levels reached are still inferior to those observed for CVD graphene on copper substrates.

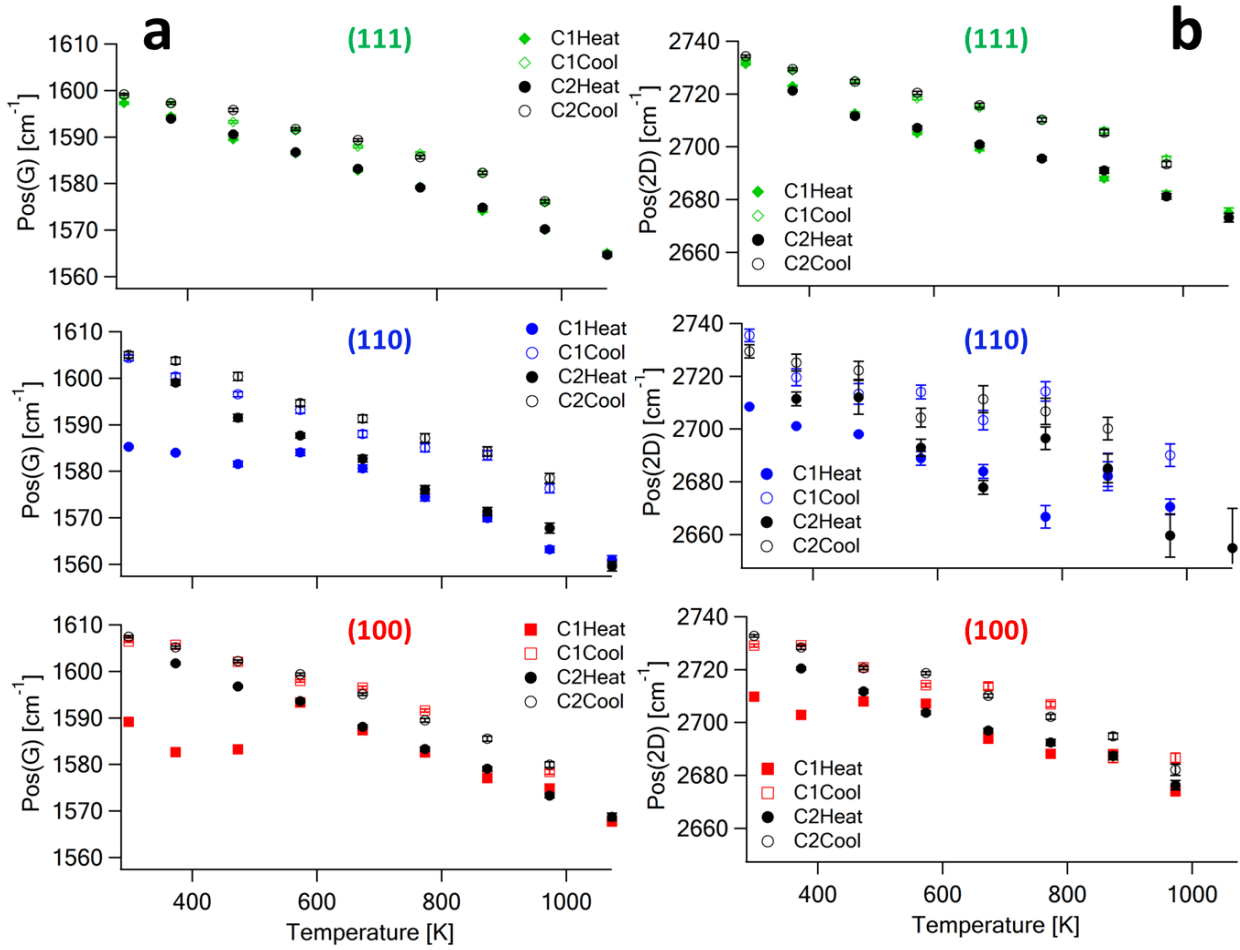

**Figure 3** Temperature dependence of (a) G band, and (b) 2D band frequencies of 1-LG grown on (100), (110), and (111) copper single crystals, during the heating (Heat) and cooling (Cool) processes for cycle 1 (C1) and cycle 2 (C2). The error bars represent the errors associated with the fitting of the Raman spectra.

Doping of graphene by charge carriers can also be monitored using Raman spectroscopy. Several effects can be observed in the Raman spectrum of doped graphene: a) upshift of the G band frequency, b) upshift or downshift of the 2D band frequency, for *p*-type or *n*-type doping, respectively, c) decrease of the 2D band intensity [13, 14]. The G and 2D bands can be found in the Raman spectra of undoped suspended graphene at ≈1580 cm$^{-1}$ and at ≈2680 cm$^{-1}$, at 2.41 eV, respectively [34]. The 2D band is dispersive, and its frequency depends on the excitation energy (e.g. ≈2700 cm$^{-1}$ at 2.54 eV) [28]. The frequency of the Raman bands can shift to higher values for supported graphene (G band ≈1587 cm$^{-1}$) [34], due to substrate induced doping and strain on graphene [35]. Furthermore, the 2D band frequency and width are sensitive to strain [17]. To investigate the effects of doping the G band position (Pos(G)) was plotted as a function of temperature, see Figure 3a, for all copper crystal orientations, during heating and cooling of both cycle 1 and cycle 2. For graphene grown on Cu(100) and Cu(110) at room temperature before heating in cycle 1, the Pos(G) is approximately 1586 and 1587 cm$^{-1}$, respectively. However, the Pos(G) reaches more than 1600 cm$^{-1}$ both after the first and after the second heating cycles for graphene on Cu(100) and Cu(110), indicating an increase of the doping level upon heating. This effect is not reversible, as the Pos(G) remains above 1600 cm$^{-1}$, at room temperature, for the following heating/cooling procedures. In contrast, for graphene grown on Cu(111) the Pos(G) is ≈ 1600 cm$^{-1}$ at room temperature for all cycles, indicating a similar level of doping before and after the heating/cooling cycles. Similar effects are observed when the 2D band frequency is plotted as a function of the temperature, see Figure 3b. For graphene grown on Cu(111) the Pos(2D) at room temperature is approximately the same (2735 cm$^{-1}$) before and after heating, while for graphene grown on Cu(100) and Cu(110) the Pos(2D) increases (from 2710 to about 2735 cm$^{-1}$) after the samples are heated and cooled in cycle 1 and cycle 2. The initial values (before heating) of Pos(G) are reproducible for all samples and similar results were obtained for samples stored in vacuum as well. Thus, samples exposed to air for scarce seconds, needed to transfer the single crystals from the furnace to the storage container, show a similar trend as described in Figure 3. The ΔPos(2D)/ΔPos(G) for graphene on Cu(111) is ~1 in the whole temperature range, which evidences the doping to be the dominant factor of the observed changes in the spectra, as opposed to strain, which would cause the ΔPos(2D)/ΔPos(G) ratio of ~2.5 [32].

The shift of the Fermi level and thereby the amount of doping can be deduced from the difference of the metal substrate and the graphene work functions [36]. The work function is an intrinsic property of the material, yet it can be modified by chemical doping [36, 37]. For instance, the work function of the copper covered with graphene is different from the pristine copper, and it depends on the separation between the copper and the graphene [36]. As the distance between the copper and the graphene layer decreases, the electronic structure of graphene is modified, thus increasing the doping level beyond the limit predicted by the work functions of graphene and copper [36]. By assuming a small mismatch between the Cu(111) surface and the graphene, their separation can be estimated as the distance between two graphene layers in graphite (d=3.3 Å) [38]. The work function of copper is also dependent on the surface orientation, and is ϕ$_{Cu(111)}$ = 4.94 eV for Cu(111) [39], whereas the work function of monolayer graphene is ϕ$_G$ = 4.57 eV [40]. The difference in work function of ϕ$_{Cu(111)}$ - ϕ$_G$ = 0.37 eV at a separation of 3.3 Å corresponds to a Fermi level shift of approximately 0.4 eV [34]. This Fermi level shift corresponds to ΔPos(G) = 12 cm$^{-1}$ (where ΔPos(G) = Pos(G)$_{doped}$-Pos(G)$_{undoped}$) [11], which is in agreement with the experimental value obtained for graphene grown on Cu(111) before the heating cycles, see Figure 3a.

The work functions of Cu(100) and Cu(110) were found to be: ϕ$_{Cu(100)}$ = 4.59 eV and ϕ$_{Cu(110)}$ = 4.48 eV, respectively. However, the weaker interaction between graphene and the Cu(100) and Cu(110) surfaces leads to a large separation of graphene-copper, which results in a very small Fermi level shift for both. Consequently, the doping from the copper is only negligible initially, resulting in a very small shift of the G peak position before the heating cycles.

When the graphene is heated in the two heating cycles, the interaction between the graphene and the Cu(100) and Cu(110) surfaces increases. The mismatch between the graphene and the Cu(100) and Cu(110) surfaces allows an even closer contact than between the graphene and the Cu(111), where the atoms repel each other since they are placed on top of each other. The distance between the graphene and the Cu(111) therefore does not change after heating, leading to a similar doping and G peak shift as before the heating ($\Delta Pos(G)=14$ cm$^{-1}$). However, the graphene on the Cu(100) and Cu(110) surfaces get in close contact and their interaction changes progressively from physisorption to chemisorption, shifting the Fermi level (or effective work function), which in turn leads to an increased doping level [36]. This can be seen in the large $\Delta Pos(G)$ at room temperature of 21 and 19 cm$^{-1}$ after the first heating cycle for the graphene grown on Cu(100) and Cu(110), respectively. The separation dependent chemical interaction has theoretically been shown to be able to shift the sign of the doping, from *p*-type to *n*-type as the separation is decreased [36] and it can also explain the loss of Raman signal from graphene on platinum, for instance, as the graphene gets so doped that it starts behaving as a metal [41]. Moreover, our results can also explain why graphene on Ir has been shown to yield a strong Raman signal or no Raman signal at all [42-44], depending on the mismatch between the graphene and the surface orientation of the substrate.

The increase of contact between the graphene layer and the copper surfaces also explains the broadening in the G band after cycle 1 (increase of 36, 5, and 4 cm$^{-1}$ for Cu(100), Cu(110) and Cu(111), respectively). The surface roughness is much higher for graphene grown on Cu(110), than on Cu(100) and Cu(111), as was previously investigated by atomic force microscopy [11]. The reduced distance between graphene and copper enhances the effects of the inhomogeneities found on the surface, leading to a broadening of the G band. The mismatch of the graphene and copper lattices is also reflected in the linear dependence of the Pos(G) on temperature (between 298 and 1073 K) in the second heating cycle, which is 0.041, 0.047 and 0.054 cm$^{-1}$K$^{-1}$ for graphene grown on Cu(111), Cu(100) and Cu(110), respectively. The higher values of graphene on Cu(100) and Cu(110) corroborate the higher adhesion of the graphene to the Cu(110) and Cu(100) surfaces. These differences in temperature shift rates closely resemble the observed variations in the case of strongly adhering graphene layer to the Si/SiO$_2$ substrate (higher values of the shift rates) in contrast to weakly adhering or suspended graphene (lower values of the shift rates) [18, 20, 45].

Another noticeable effect is the behavior of the G and 2D bands during the heating and cooling processes for temperatures higher than 500 K. During the cooling procedure the position of the Raman bands is higher than during the heating process at each particular temperature, leading to a hysteresis effect. A similar hysteresis effect has been reported for graphene grown on iridium substrates, as a consequence of wrinkle formation during the cooling processes [16]. The Pos(G) is sensitive to strain in the graphene layer as graphene and copper have thermal expansion coefficients of opposite signs. Consequently, the Pos(G) changes differently as the temperature increases or decreases due to strain effects, e.g. wrinkle formation, eventually causing the hysteresis effect observed in Figure 3a.

Immediately after the growth in the CVD furnace, the graphene undergoes a cooling process under similar conditions (pressure and temperature) as in the case of the cooling cycle investigated here. Several interesting aspects can therefore be concluded about this important, and often neglected, part of the graphene synthesis. The graphene is in close contact with the copper during the growth. Afterwards however, as soon as the samples are cooled down and/or exposed to air, the graphene can instantly be detached from the copper, depending on the copper surface orientation. Graphene grown on Cu(100) and Cu(110) becomes detached upon cooling/air exposure, as shown by the reduced doping, while graphene grown on Cu(111) remains in close contact with the substrate as evidenced by the constant doping level. This indicates that transfer of graphene grown on Cu(100) and Cu(110) onto other substrates should be easier than of graphene grown on Cu(111). The strong interaction between graphene and a metallic surface with similar lattice orientation also explains the lack of Raman signal on such substrates. Hence, the detachment of graphene from the metallic substrate plays an important role for the Raman measurements as well.

## 4. Conclusion

The investigation of as-grown graphene on single crystals of copper has shown that graphene is doped and its doping level depends on the surface orientation of the crystal as follows: Cu(111) > Cu(100) > Cu(110). This order can be tuned or even reversed upon a subsequent heating/cooling cycle in inert atmosphere: Cu(100) ≈ Cu(110) > Cu(111). When graphene is grown on copper, a charge transfer between the copper and the graphene levels occurs and the graphene becomes doped. Besides the charge transfer, a contribution from metal-graphene chemical interaction also exists, and this contribution increases when the distance between the two materials decreases, such as after the materials are heated in an inert atmosphere. Consequently, heating graphene on copper leads to doping levels beyond what is estimated by the difference between the copper and graphene work functions.

Our results also indicate that graphene is delaminated from the Cu(100) and Cu(110) substrate during the cooling/air exposure after the growth, while graphene remain in close contact with Cu(111) substrate. This implies some difficulties, which can be experienced during the transfer process from polycrystalline copper foils. Nevertheless, a subsequent heating can improve the graphene substrate interactions also for Cu(100) and Cu(110) surface orientations. Our findings are also important for understanding the effects of temperature variations on the graphene layers, such as the ones used during the CVD process, representing a step forward towards the production and transfer of high-quality and large-area graphene sheets with pre-defined properties.

**Acknowledgements**

The authors acknowledge the support of MSMT ERC-CZ project (LL 1301).